\documentclass[twocolumn,showpacs,aps,prl,letterpaper]{revtex4}
\usepackage{graphicx}
\usepackage{dcolumn}
\usepackage{amsmath}
\usepackage{epsfig}
\long\def\inst#1{\par\nobreak\kern 4pt\nobreak
    {\itshape #1}\par\vskip 10pt plus 3pt minus 3pt}
\RequirePackage{xspace}
\usepackage{relsize}

\def\babar{\mbox{\slshape B\kern-0.1em{\smaller A}\kern-0.1em
    B\kern-0.1em{\smaller A\kern-0.2em R}}}
\def\Kbar    {\kern 0.18em\overline{\kern -0.18em K}{}\xspace}

\def\KK      {\ensuremath{K\Kbar}\xspace}
\def\Kz      {\ensuremath{K^0}\xspace}
\def\Kzb     {\ensuremath{\Kbar^0}\xspace}
\def\KzKzb   {\ensuremath{\Kz {\kern -0.16em \Kzb}}\xspace}

\def\Ks     {\ensuremath{K_S}\xspace}
\def\Kl     {\ensuremath{K_L}\xspace}
\def\KsKs   {\ensuremath{\Ks {\kern -0.16em \Ks}}\xspace}
\def\KlKl   {\ensuremath{\Kl {\kern -0.16em \Kl}}\xspace}
\def\KsKl   {\ensuremath{\Ks {\kern -0.16em \Kl}}\xspace}
\def\KlKs   {\ensuremath{\Kl {\kern -0.16em \Ks}}\xspace}
\def\Dbar    {\kern 0.18em\overline{\kern -0.18em D}{}\xspace}
\def\Bbar    {\kern 0.18em\overline{\kern -0.18em B}{}\xspace}

\def\Bz      {\ensuremath{B^0}\xspace}
\def\Bzb     {\ensuremath{\Bbar^0}\xspace}
\def\BzBzb   {\ensuremath{\Bz {\kern -0.16em \Bzb}}\xspace}
\def\Bu      {\ensuremath{B^+}\xspace}
\def\Bub     {\ensuremath{B^-}\xspace}

\def\BpBm    {\ensuremath{\Bu {\kern -0.16em \Bub}}\xspace}

\newcommand{\optbar}[1]{\shortstack{{\tiny (\rule[.4ex]{1em}{.1mm})}
  \\ [-.7ex] $#1$}}
\def\BorBbar    {\kern 0.18em\optbar{\kern -0.18em B}{}\xspace}
\def\DorDbar    {\kern 0.18em\optbar{\kern -0.18em D}{}\xspace}
\def\KorKbar    {\kern 0.18em\optbar{\kern -0.18em K}{}\xspace}
\def\CP                {\ensuremath{C\!P}\xspace}
\def\pep2{PEP-II}
\mathchardef\Upsilon="7107
\def\Y#1S{\ensuremath{\Upsilon{(#1S)}}\xspace}

\begin{document}

\title{
\large \bfseries \boldmath
Clean Prediction of \CP Violating Processes $\psi$, $\phi$ and
$\Upsilon(1S)$ Decay to \KsKs and \KlKl }

\author{Haibo Li}\email{lihb@mail.ihep.ac.cn}
\author{Maozhi Yang}\email{yangmz@mail.ihep.ac.cn}
\affiliation{ CCAST (World Laboratory), P.O.Box 8730, Beijing 100080, China  \\
 Institute of High Energy Physics, P.O.Box 918, Beijing  100049, China }


\date{\today}


\begin{abstract}
The ratio of $\KsKs$ ($\KlKl$) and  $\KsKl$
production rates is calculated by considering $\Kz - \Kzb$ oscillation
in $J/\psi \rightarrow \KzKzb$ decay.
The theoretical uncertainty due to strong interaction in $J/\psi$
decay is completely canceled in the ratio,  therefore,
 the absolute branching fractions
of the \CP violating processes of $J/\psi \rightarrow \KsKs $ and $\KlKl$ can be
cleanly and model-independently determined in case that $J/\psi \rightarrow
\KsKl$ decay is precisely measured.
In the future $\tau$-Charm factory, the expected
\CP violating process of $J/\psi \rightarrow \KsKs $ should be reached.
It is important to measure  $J/\psi$ to $\KsKs$ and \KsKl decays simultaneously,
so that many systematic errors will be canceled.
More precise measurements are suggested to
examine the predicted isospin relation in $J/\psi \rightarrow \KK$
decays. All results can be extended to decays of other vector quarkonia, $\phi$, $\psi(2S)$ and $\Upsilon(1S)$ and
so on.

\end{abstract}


\maketitle


In the Standard Model (SM), \CP violation arises from
an irreducible weak phase in the Cabibbo-Kobayashi-Maskawa (CKM)
quark-mixing matrix~\cite{Kobayashi}.  \CP violation has been established
in both $K$ and $B$ systems. Currently,  all experimental measurements are consistent
with the CKM picture of \CP violation, and the CKM is, very
likely, the dominant source of \CP violation in low energy flavor-changing
processes~\cite{nir}.  However, the surprising point is that the CKM mechanism for \CP
violation fails to account for the baryogenesis~\cite{baryon}. It
is crucial to probe \CP violation in various reactions, to see the
correlations between different processes and probe the source of \CP
violation.

In this Letter, we consider the possible \CP asymmetric observation
in $J/\psi \rightarrow \KzKzb$ decay, in which \KsKs and \KlKl
pairs can be formed in addition to $\KsKl$.
Within the SM, the possible \CP violating decay
processes of $J/\psi \rightarrow \KsKs$ and $\KlKl$ are due to $\Kz
-\Kzb$ oscillation, here we assume that possible strong multiquark
effects that involve seaquarks play no role in $J/\psi\to \KsKs$,
$\KlKl$, and $\KsKl$ decays~\cite{voloshin}. The $J/\psi$ decays will provide
another opportunity to understand the source of \CP violation. The
amplitude for $J/\psi$ decaying to $\KzKzb$ is $\langle
\KzKzb|H|J/\psi \rangle$, and the $\KzKzb$ pair system is in a
state with charge parity $C= -1$,  which can be defined as
\begin{equation}
|\KzKzb\rangle^{C=-1} = \frac{1}{\sqrt{2}} \left [
|\Kz \rangle |\Kzb\rangle -
|\Kzb\rangle |\Kz\rangle\right ].
\label{eq:k0k0}
\end{equation}
Although there is weak current contribution in $J/\psi \rightarrow
\KzKzb$ decay, which may not conserve charge parity, the $\KzKzb$
pair can not be in a state with $C= +1$. The reason is that the
relative orbital angular momentum of $\KzKzb$ pair must be $l=1$
because of angular momentum conservation. A boson-pair with $l=1$
must be in an anti-symmetric state, the anti-symmetric state of
particle-anti-particle pair must be in a state with $C= -1$. This
conclusion can also be illustrated by a direct calculation.

  For explicitness, let us denote the particle state $|\Kz\rangle$ with
momentum $p_1$ as $|\Kz (p_1)\rangle$, and $|\Kzb\rangle$ with
momentum $p_2$ as $|\Kzb (p_2)\rangle$. The general structure of
the effective Hamitonian for $J/\psi \rightarrow \KzKzb$ decay is
\begin{equation}
H=C_W \sum_{i=d,s}\bar{q}_i\gamma^{\mu}(a-b\gamma_5)q_i
\bar{c}\gamma_{\mu}(a' -b'\gamma_5)c ,
\end{equation}
where $q_i$ ($i=d,s$) and $c$ denote the quarks $d$, $s$ and $c$,
$C_W$ is the Wilson coefficient, and $a$, $b$, $a'$, $b'$ are the
relevant coefficients for the vector and axial-vector currents.
Then the amplitude for $J/\psi$ decaying into a state $|\Kz
(p_1)\Kzb (p_2)\rangle$ is
\begin{eqnarray}\label{e3}
&&\langle \Kz (p_1)\Kzb (p_2)|H|J/\psi\rangle \nonumber
\\
&=&C_W a' \langle \Kz (p_1)\Kzb
(p_2)|\sum_{i=d,s}\bar{q}_i\gamma^{\mu}(a-b\gamma_5)q_i |0\rangle  \nonumber
\\
&& \cdot \langle 0 |\bar{c}\gamma_{\mu}c|J/\psi\rangle,
\end{eqnarray}
where $\langle 0 |\bar{c}\gamma^{\mu}\gamma_5 c|J/\psi\rangle =0$
has been used. From the Lorentz structure of the above matrix
element, we have the following decomposition
\begin{eqnarray}\label{e4}
&&\langle \Kz (p_1)\Kzb
(p_2)|\sum_{i=d,s}\bar{q}_i\gamma^{\mu}(a-b\gamma_5)q_i |0\rangle
\nonumber \\
&=&F^+(p_1+p_2)^{\mu}+F^-(p_1-p_2)^{\mu},
\end{eqnarray}
where $F^+$ and $F^-$ are Lorentz invariant form factors. For the
vector current induced vacuum-$J/\psi$ matrix element, there is
the common decomposition
\begin{equation}\label{e5}
\langle 0
|\bar{c}\gamma_{\mu}c|J/\psi\rangle=f_{J/\psi}m_{J/\psi}\epsilon_{\mu},
\end{equation}
where $f_{J/\psi}$ is the decay constant, and $\epsilon_{\mu}$ the
polarization vector of $J/\psi$.

 Substitute Eqs.(\ref{e4}) and (\ref{e5}) into Eq.(\ref{e3}), and
use $\epsilon \cdot (p_1+p_2)=\epsilon \cdot p=0$, where $p$ is
the momentum of $J/\psi$, we can obtain
\begin{equation}
\langle \Kz (p_1)\Kzb
(p_2)|H|J/\psi\rangle=C_Wa'f_{J/\psi}m_{J/\psi}F^-\epsilon \cdot
(p_1-p_2).
\end{equation}
Exchange $p_1$ and $p_2$, we can get another amplitude
\begin{equation}
\langle \Kzb (p_1)\Kz
(p_2)|H|J/\psi\rangle=C_Wa'f_{J/\psi}m_{J/\psi}F^-\epsilon \cdot
(p_2-p_1).
\end{equation}
The above two equations directly give
\begin{equation}
\langle \Kz (p_1)\Kzb (p_2)+\Kzb (p_1)\Kz (p_2)|H|J/\psi\rangle=0,
\end{equation}
which shows that the amplitude for $J/\psi$ decaying into $\KzKzb$
pair with charge parity $C=+1$ is zero. Therefore, only state with
$C=-1$ can be produced in $J/\psi \rightarrow \KzKzb$ decay.

Note that Eq.(\ref{e3}) is based on the factorization
approximation. However, the conclusion that the $\KzKzb$ pair
produced from a vector meson decay must be in $C=-1$ state does
not depend on the factorization approximation, it can be drawn
directly from the momentum conservation. The above computation
should be only viewed as a complementary illustration rather than demonstration.

Next we shall analyze the time-evolution of $\KzKzb$ system
produced in $J/\psi$ decay.

The weak eigenstates of $\Kz - \Kzb$ system are $|\Ks\rangle = p
|\Kz\rangle+q|\Kzb\rangle$ and $|\Kl\rangle = p |\Kz\rangle
-q|\Kzb\rangle$ with eigenvalues $\mu_S = m_S -
\displaystyle\frac{i}{2} \Gamma_S$ and $\mu_L = m_L -
\displaystyle\frac{i}{2} \Gamma_L$, respectively, where the $m_S$
and $\Gamma_S$ ($m_L$ and $\Gamma_L$) are the mass and width
of \Ks (\Kl) meson. Following the $J/\psi \rightarrow \KzKzb$
decay, the $\Kz$ and $\Kzb$ will go separately and the
time-evolution of the particle states $|\Kz(t)\rangle$ and
$|\Kzb(t)\rangle$ are given by $|\Kz(t)\rangle =
\displaystyle\frac{1}{2p} \left ( e^{-i\mu_S t} |\Ks \rangle +
e^{-i\mu_L t}|\Kl \rangle \right )$ and $|\Kzb(t) \rangle =
\displaystyle\frac{1}{2q} \left ( e^{-i\mu_S t} |\Ks \rangle -
e^{-i\mu_L t}|\Kl \rangle \right )$, respectively. Then the
time-evolution of $\Kz -\Kzb$ system with $C = -1$ is
\begin{eqnarray}
|\KzKzb (t_1, t_2) \rangle^{C=-1}
= ~~~~~~~~~~~~~~~~~~~~~~ \cr
\frac{1}{\sqrt{2}} \left[
|\Kz(t_1)
\rangle|\Kzb(t_2) \rangle - |\Kzb(t_1) \rangle |\Kz(t_2) \rangle \right]
~\cr
= \frac{1}{2\sqrt{2}pq} \left [g_{LS} |\Kl \rangle|\Ks \rangle -
    g_{SL} |\Ks \rangle|\Kl \rangle \right ] ,
 \label{eq:k0k0_time}
\end{eqnarray}
where $g_{LS} = \displaystyle e^{-i\mu_L t_1 - i \mu_S t_2}$ and $g_{SL} =
\displaystyle e^{-i\mu_S t_1 - i \mu_L t_2}$.  Since the
states $|\Ks \rangle$ and $|\Kl \rangle$ are un-orthogonal, we have
$\langle \Ks|\Kl \rangle
= \langle \Kl|\Ks \rangle = |p|^2 - |q|^2$ and
$\langle \Ks|\Ks \rangle = \langle \Kl|\Kl \rangle = 1$.
Then the amplitudes to find $\KsKs$, $\KsKl$, $\KlKs$ and $\KlKl$
 pairs are
\begin{eqnarray}
A_1(t_1, t_2) \equiv \langle \KsKs|\KzKzb (t_1, t_2) \rangle^{C=-1}
 \cr
=\frac{1}{2\sqrt{2}pq} \left [(|p|^2 - |q|^2)( g_{LS} -
    g_{SL}) \right ],
 \label{eq:ksks}
\end{eqnarray}
\begin{eqnarray}
 A_2(t_1, t_2) \equiv \langle \KsKl|\KzKzb (t_1, t_2) \rangle^{C=-1}
 \cr
=\frac{1}{2\sqrt{2}pq} \left [ g_{LS} - (|p|^2 - |q|^2)^2g_{SL}
\right ],
 \label{eq:kskl}
\end{eqnarray}
\begin{eqnarray}
A_3(t_1, t_2) \equiv \langle \KlKs|\KzKzb (t_1, t_2) \rangle^{C=-1}
 \cr
=\frac{1}{2\sqrt{2}pq} \left [(|p|^2 - |q|^2)^2g_{LS} -
    g_{SL} \right ],
 \label{eq:klks}
\end{eqnarray}
\begin{eqnarray}
A_4(t_1, t_2) \equiv \langle \KlKl|\KzKzb (t_1, t_2)\rangle^{C=-1}
 \cr
=\frac{1}{2\sqrt{2}pq} \left [(|p|^2 - |q|^2)( g_{LS} -
    g_{SL}) \right ].
 \label{eq:klkl}
\end{eqnarray}
Therefore, one can find  $A_1(t_1, t_2) = A_4(t_1, t_2)$ and
$A_3(t_1, t_2) = - A_2(t_2, t_1)$. If \CP is conserved, i.e.,
$|q/p| = 1$, or the two particles are observed at the same time,
namely, $t_1 = t_2$, then $A_1(t_1, t_2) = A_4(t_1, t_2) = 0$ for
$\KsKs$ and $\KlKl$ cases because Bose-Einstein statistics
prevents two identical  bosons from being in an antisymmetric
state. In other words,  only $\Ks$ and $\Kl$ can be seen at the
same time in $J/\psi \rightarrow \KzKzb$ decay.

Squaring the amplitudes $A_i(t_1,t_2)$'s, one can get the
time-dependent possibilities to find $\KsKs$, $\KsKl$ and $\KlKl$
pairs
\begin{eqnarray}
&&\frac{d^2{\cal P} [\Ks(t_1), \Ks(t_2)]}{dt_1dt_2}\equiv {\cal
N}_f|A_1 (t_1,
t_2)|^2\nonumber\\
&& ={\cal N}_f\frac{(|p|^2 -
|q|^2)^2}{4|pq|^2}e^{-\Gamma(t_1+t_2)}[cosh(y\Gamma(t_2-t_1)\nonumber\\
&&~-cos(x\Gamma(t_2-t_1)],
\end{eqnarray}
\begin{eqnarray}
&&\frac{d^2{\cal P} [\Kl(t_1), \Ks(t_2)]}{dt_1dt_2}\equiv {\cal
N}_f|A_2 (t_1,
t_2)|^2\nonumber\\
& =&{\cal N}_f
\frac{1}{8|pq|^2}e^{-\Gamma(t_1+t_2)}[e^{y\Gamma(t_2-t_1)}-2(|p|^2-|q|^2)^2\nonumber\\
&\times& cos(x\Gamma(t_2-t_1))
+(|p|^2-|q|^2)^4e^{-y\Gamma(t_2-t_1)}],
\end{eqnarray}
\begin{eqnarray}
&&\frac{d^2{\cal P} [\Ks(t_1), \Kl(t_2)]}{dt_1dt_2}\equiv {\cal
N}_f|A_3 (t_1,
t_2)|^2\nonumber\\
& =&{\cal N}_f|A_2 (t_2, t_1)|^2,
\end{eqnarray}
\begin{eqnarray}
&&\frac{d^2{\cal P} [\Kl(t_1), \Kl(t_2)]}{dt_1dt_2}\equiv {\cal
N}_f|A_4 (t_1,
t_2)|^2\nonumber\\
& =&{\cal N}_f|A_1 (t_1, t_2)|^2,
\end{eqnarray}
where ${\cal N}_f$ is a common normalization factor, $\Gamma =
\displaystyle\frac{\Gamma_S+\Gamma_L}{2}$, $x =\displaystyle
\frac{\Delta m}{\Gamma}$ and $y = \displaystyle\frac{\Delta
\Gamma}{2\Gamma}$ ($\Delta m$ is the mass difference of $\Kl$ and
$\Ks$, i.e., $\Delta m = m_L - m_S$, while $\Delta \Gamma =
\Gamma_L - \Gamma_S$ is the width difference).

The time-integrated possibilities to observe $\KsKs$, $\KsKl$ and
$\KlKl$ pairs, which are normalized by the widths of $\Ks$ and
$\Kl$, i.e., ${\cal N}_f=\Gamma_S \Gamma_L$, are
\begin{eqnarray}
{\cal P} (\Ks, \Ks) \equiv \Gamma_S \Gamma_L \int^{\infty}_0
dt_2 \int^{\infty}_0 dt_1 |A_1 (t_1, t_2)|^2
\cr
= \frac{ \Gamma_S \Gamma_L}{\Gamma^2} \frac{(|p|^2 -
|q|^2)^2}{4|pq|^2} \left( \frac{1}{1-y^2} - \frac{1}{1+x^2} \right),
\label{eq:ksks_p}
\end{eqnarray}
\begin{eqnarray}
&&{\cal P}(\Ks, \Kl) \equiv \nonumber \\
&& \Gamma_S \Gamma_L \int^{\infty}_0
dt_2 \int^{\infty}_0 dt_1 (|A_2 (t_1, t_2)|^2 + |A_3 (t_1, t_2)|^2 \nonumber \\
& = & \frac{ \Gamma_S \Gamma_L}{\Gamma^2} \frac{1}{4|pq|^2}\times \nonumber \\
& & \left ( \frac{1}{1-y^2}
- \frac{2(|p|^2 - |q|^2)^2}{1+x^2}+\frac{(|p|^2-|q|^2)^4}{1-y^2}\right ),
\label{eq:kskl_p}
\end{eqnarray}
\begin{eqnarray}
{\cal P}(\Kl, \Kl) &\equiv& \Gamma_S \Gamma_L \int^{\infty}_0
dt_2 \int^{\infty}_0 dt_1 |A_4 (t_1, t_2)|^2 \nonumber \\
 &=& {\cal P} (\Ks, \Ks).
\label{eq:klkl_p}
\end{eqnarray}

The final amplitude squared for the
decay process $J/\psi \rightarrow K_1 K_2$, where $K_1 K_2$ can be $\KsKs$,
$\KsKl$ or $\KlKl$, which is from the time-evolution of $\Kz -
\Kzb$ system, is
\begin{equation}
|A(J/\psi \rightarrow K_1 K_2)|^2 = {\cal P}(K_1, K_2) |\langle \KzKzb|H|J/\psi \rangle |^2,
\label{eq:psi-kk}
\end{equation}
where ${\cal P} (K_1, K_2)$ can be obtained from
Eqs.~(\ref{eq:ksks_p}),~(\ref{eq:kskl_p}) and~(\ref{eq:klkl_p}).
Finally, the partial widths of the $J/\psi \rightarrow \KsKs$,
$\KsKl$ and $\KlKl$ are
\begin{eqnarray}
\Gamma(\psi \rightarrow \KsKs) = {\cal P}(\Ks, \Ks)\cdot C \cdot
|\langle \KzKzb|H|\psi \rangle |^2,
\label{eq:psi-ksks}
\end{eqnarray}
\begin{eqnarray}
\Gamma (\psi \rightarrow \KsKl) = {\cal P}(\Ks, \Kl)\cdot C \cdot |\langle
\KzKzb|H|\psi \rangle |^2,
\label{eq:psi-kskl}
\end{eqnarray}
\begin{eqnarray}
\Gamma (\psi \rightarrow \KlKl) =  {\cal P}(\Kl, \Kl)\cdot C \cdot
|\langle \KzKzb|H|\psi \rangle |^2,
\label{eq:psi-klkl}
\end{eqnarray}
where $C = \displaystyle \frac{1}{3} \frac{1}{8\pi} \frac{|\bf P|}{m_{J/\psi}^2}$
is the phase space factor,
$m_{J/\psi}$ is the $J/\psi$ mass and ${\bf P}$ is the three-momentum of the final particles in the rest
frame of $J/\psi$. Combining Eqs.~(\ref{eq:klkl_p}),~(\ref{eq:psi-ksks})
and~(\ref{eq:psi-klkl}), one can get $\Gamma (J/\psi \rightarrow \KsKs) =
\Gamma (J/\psi \rightarrow \KlKl) $.  The ratio $R_{SS}$
($R_{LL}$) of $\KsKs$ ($\KlKl$) and $\KsKl$
production rates is obtained
\begin{eqnarray}
R_{SS} \equiv  \frac{\Gamma(J/\psi \rightarrow \KsKs)}{\Gamma(J/\psi \rightarrow
\KsKl)}
~~~\cr
=(|p|^2 - |q|^2)^2 \frac{x^2+y^2}{1+x^2},
\label{eq:rss}
\end{eqnarray}
and
\begin{eqnarray}
R_{LL} \equiv \frac{\Gamma(J/\psi \rightarrow \KlKl)}{\Gamma(J/\psi \rightarrow
\KsKl)} = R_{SS}.
\label{eq:rll}
\end{eqnarray}
In the ratios, the phase space factor $C$ and the strong
matrix element squared $|\langle \KzKzb|H|J/\psi \rangle |^2$ are
completely canceled, which ensures that the ratios are completely free
from uncertainty caused by strong
interaction in $J/\psi \rightarrow \KzKzb$ decay.
The expected value of the ratio, $R_{SS}$ ($R_{LL}$), can be determined
by using experimental measured values of $x$, $y$ and $|p|^2 - |q|^2$, where
$|p|^2 - |q|^2$ is related to the \CP asymmetry parameter $\delta_L$ in
semileptonic $\Kl$ decay
\begin{eqnarray}
 \delta_L  &\equiv &  \frac{\Gamma(\Kl \rightarrow l^+ \nu \pi^-) -
\Gamma(\Kl \rightarrow l^- \nu \pi^+) } {
\Gamma(\Kl \rightarrow l^+ \nu \pi^- )+
\Gamma(\Kl \rightarrow l^- \nu \pi^+)} \nonumber\\
&=& |p|^2 - |q|^2. \label{eq:deta}
\end{eqnarray}
One can further express $R_{SS}$ ($R_{LL}$) completely in terms of
mixing and $CP$ asymmetry parameters in kaon decays, by combining
Eqs.(\ref{eq:rss}), (\ref{eq:rll}) and (\ref{eq:deta})
\begin{eqnarray}
R_{SS} =R_{LL}=(\delta_L)^2 \frac{x^2+y^2}{1+x^2}, \label{eq:sl}
\end{eqnarray}
where the experimental result of $\delta_L$ is $(3.27 \pm 0.12)
\times 10^{-3}$~\cite{pdg2004}. From the measured values of
mass difference and widths of $\Kl$ and $\Ks$ \cite{pdg2004}, one
can get $x = 0.946$ and $y = -0.997$. The uncertainties on $x$ and
$y$ are negligible. Therefore,
$\displaystyle\frac{x^2+y^2}{1+x^2}\approx 1$ and $R_{SS}
=R_{LL}\approx (\delta_L)^2$. The dominated
uncertainty of $R_{SS}$ ($R_{LL}$) is from the error of the measured
value of $\delta_L$. The values of the ratios $R_{SS}$ and $R_{LL}$ can
be obtained
\begin{eqnarray}
R_{SS} = R_{LL} = (10.66 \pm 0.78 ) \times 10^{-6},
\label{eq:rvalue}
\end{eqnarray}
which are model-independent and theoretically clean. The total
uncertainty of $R_{SS}$ ($R_{LL}$) is about 7\%, and can be
improved in the future if more precise measurement of \CP
violation parameter in $\Kl \rightarrow l^{\pm} \nu \pi^{\mp} $
decays is obtained. In general, one can measure
$J/\psi$ to $\KsKs$ and $\KsKl$ decays simultaneously in the
future $\tau$-Charm factory, so that some of the systematic errors
can be canceled in the ratio.
It is very interesting that the ratio can be extended
to other quarkonia which can decay into $\KK$ final states, for example,
in $\phi$, $\psi(2S)$ or $\Upsilon(1S) \rightarrow \KzKzb$ decays,
we have
\begin{eqnarray}
R_{SS}&\equiv& \frac{\Gamma(J/\psi \rightarrow \KsKs)}{\Gamma(J/\psi \rightarrow
\KsKl)} = \frac{\Gamma(\psi(2S) \rightarrow \KsKs)}{\Gamma(\psi(2S) \rightarrow
\KsKl)} \nonumber\\
&=& \frac{\Gamma(\phi \rightarrow \KsKs)}{\Gamma(\phi \rightarrow \KsKl)}
= \frac{\Gamma(\Upsilon(1S) \rightarrow \KsKs)}{\Gamma(\Upsilon(1S) \rightarrow
\KsKl)}.
\label{eq:rvalue_extend}
\end{eqnarray}

With current available experimental data, such as, ${\cal B}
(J/\psi \rightarrow \KsKl) = (1.82 \pm 0.04 \pm 0.13) \times
10^{-4}$~\cite{bes1}, ${\cal B} (\psi(2S) \rightarrow \KsKl) =
(5.24 \pm 0.47 \pm 0.48) \times 10^{-5}$~\cite{bes2} and ${\cal B}
(\phi \rightarrow \KsKl) = (33.7 \pm 0.5)\%$~\cite{pdg2004}, the
branching fractions of \CP violating decay processes of $J/\psi$,
$\psi(2S)$ and $\phi \rightarrow \KsKs (\KlKl)$ can be extracted
as following
\begin{eqnarray}
{\cal B}(J/\psi \rightarrow \KsKs) = (1.94 \pm 0.20)  \times 10^{-9}, \nonumber\\
{\cal B}(\psi(2S) \rightarrow \KsKs) = (0.56 \pm 0.08 ) \times 10^{-9},\\
{\cal B}(\phi \rightarrow \KsKs) = (3.59 \pm 0.27) \times 10^{-6}, \nonumber
\label{eq:predict}
\end{eqnarray}
which are the first model-independent predictions in rare
quarkonium decays. With more precise measurements of \KsKl decays, the
above errors can be reduced further.  It is interesting that the $\phi
\rightarrow \KsKs$ decay can be even reached at the current KLOE experiment at
the DA$\phi$NE accelerator.

The \CP violating decay processes of $J/\psi \rightarrow \KsKs$
and $\psi(2S) \rightarrow \KsKs$ had been searched for by BESII
Collaboration. The upper limits on the branching fractions at 95\%
C.L. are set: ${\cal B} (J/\psi \rightarrow \KsKs) < 1.0 \times
10^{-6}$ and ${\cal B} (\psi(2S) \rightarrow \KsKs) < 4.6 \times
10^{-6}$, respectively \cite{bes3}.  The current bounds of the
production rates are beyond the sensitivity for testing $R_{SS}$.
However, the BESIII experiment will start to take data in the
middle of 2007. About $10 \times 10^{9}$ $J/\psi$ and $3 \times
10^{9}$ $\psi(2S)$ data samples can be collected per year's
running according to the designed luminosity of BEPCII in
Beijing~\cite{besiii}. Thus, both $J/\psi \rightarrow \KsKs$ and
$\psi(2S) \rightarrow \KsKs$ will be reached with data taking in a
few years at BEPCII. The $\phi \rightarrow \KsKs$ will also be
easily accessible at the future DA$\phi$NEII Frascati
$\phi$-factory at which the designed luminosity is about $1.0
\times 10^{34}$ cm$^{-2}$s$^{-1}$~\cite{fuko},
 and the signal is clean and free from backgrounds because it is just near the
$\KK$ threshold.  $\Upsilon(1S) \rightarrow \KsKs$ and
$\KsKl$ can be studied at B-factory~\cite{fuko}, for example, with the current luminosity at KEK-B,
about $4.0 \times 10^9$ $\Upsilon(1S)$ events can be collected with one year's running.  It
will be very interesting to collect more data at the future Super-B factory to
test the $R_{SS}$.

Due to isospin symmetry, we have $\langle \KzKzb | H|J/\psi\rangle =
\langle K^+K^-|H|J/\psi \rangle$, then, it is straightforward to obtain
the following relation by neglecting the phase space difference
\begin{eqnarray}
{\cal B}(J/\psi \rightarrow \KsKl) & = & {\cal A} \cdot  {\cal B} (J/\psi \rightarrow K^+
K^-) \nonumber \\
 & \cong & {\cal B} (J/\psi \rightarrow K^+ K^-),
\label{eq:isospin}
\end{eqnarray}
where ${\cal A}$ is the correction factor due to $\Kz-\Kzb$ mixing, which
can be derived from Eqs.~(\ref{eq:kskl_p}) and~(\ref{eq:psi-kskl})
\begin{eqnarray}
{\cal A} &=& \frac{\Gamma_S \Gamma_L}{\Gamma^2} \frac{1}{4|pq|^2}
 \{ \frac{1}{1-y^2}- \nonumber \\
& & \frac{2(|p|^2-|q|^2)^2}{1+x^2} + \frac{(|p|^2-|q|^2)^4}{1-y^2} \} \nonumber \\
& = & 0.972,
\label{eq:factor}
\end{eqnarray}
where we have used the relation $\displaystyle\frac{\Gamma_S \Gamma_L}{\Gamma^2}\frac{1}{1-y^2} = 1$.
The uncertainty on ${\cal A}$ is tiny, at the order of $10^{-5}$.
The correction factor ${\cal A}$ is slightly smaller than 1, which is caused by $\Kz-\Kzb$ mixing.

Eqs.(\ref{eq:isospin}) and (\ref{eq:factor}) show that the ratio
${\cal A}=\displaystyle\frac{{\cal B}(J/\psi \rightarrow
\KsKl)}{{\cal B} (J/\psi \rightarrow K^+ K^-)}\neq 1$ even if
isospin symmetry is an exact symmetry. However isospin symmetry is
only an approximate symmetry, it is usually violated by a few
percent level. Electromagnetic effects may be even particularly
important for the the $K^+K^-$ final state. Therefore, the true
value of the factor ${\cal A}$ may be even largely different from
unity, which should be the total effects of $\Kz-\Kzb$ mixing and
isospin violation. The isospin violating effect must be
taken into account when comparing the experimental value of the ratio
$\displaystyle {\cal A} = \frac{{\cal B}(J/\psi \rightarrow \KsKl)}{{\cal B}
(J/\psi \rightarrow K^+ K^-)}$ with the theoretical prediction.
However further study of the isospin violating effect is beyond the scope of this paper.

One more remark is in the following. Soft photons can be emitted from
the initial and final states in vector quarkonia to $K\overline{K}$ decays.
The radiation of the soft photons in the decays allows the
$\KzKzb$ in a $C=+1$ state. Such a process with a soft photon in
the $\KsKs$ or $\KlKl$ final state is not $CP$-violating.
The detection of the soft photons depends on the sensitivity of
the detectors. Therefore the soft-photon-radiation process
is an experimental background for the test of the $R_{SS}$ and
$R_{LL}$ predictions. This background should be subtracted in
experiment.

 In conclusion, we have studied the \CP violating decay processes of
$J/\psi$, $\psi(2S)$, $\phi$ and $\Upsilon(1S)$ quarkonia to $\KsKs$ and $\KlKl$.
The ratio $R_{SS}$ ($R_{LL}$) of
$\KsKs$ ($\KlKl$) and $\KsKl$ production rates has
been constructed in a model-independent and theoretically clean
way. Simultaneous measurements of vector quarkonium decays to both
$\KsKs$ and $\KsKl$ pairs are suggested at higher luminosity
$e^+e^-$ machines, so that many systematic errors can be
canceled. With the current experimental information, the absolute
branching fractions of the \CP violating processes are firstly
predicted.  The isospin relation is obtained by considering
$\Kz - \Kzb$ mixing effect.

This work is supported in part by the National Natural Science Foundation
of China under contracts Nos. 10205017, 10575108, and the Knowledge Innovation Project of
CAS under contract Nos. U-612 and U-530 (IHEP).





\begin{thebibliography}{99}

\bibitem{Kobayashi}
N.~Cabibbo, Phys.\  Rev.\ Lett. {\bf 10}, 531 (1963);
M.~Kobayashi and T.~Maskawa, Prog. Theor. Phys. {\bf 49}, 652 (1973).

\bibitem{nir}
Yosef Nir, hep-ph/0510413.

\bibitem{baryon}
A.~D.~Sakharov, JETP Lett. {\bf 5}, 24 (1967); G.~R.~Farrar and
M.~E.~Shaposhnikov, Phys. Rev. D {\bf 50}, 774 (1994); P.~Huet and
E.~Sather, Phys. Rev. D {\bf 51}, 379 (1995).

\bibitem{voloshin}
M.~B.~Voloshin, Phys.~Rev.~{\bf D71}, 114003 (2005).
\bibitem{pdg2004}
 S. Eidelman {\it et al.}, Phys. Lett. B {\bf 592}, 1 (2004).
\bibitem{bes1}
J.Z.~Bai {\it et al.},  Phys. Rev. D {\bf 69}, 012003 (2004).
\bibitem{bes2}
J.Z.~Bai {\it et al.},  Phys. Rev. Lett. {\bf 92}, 052001 (2004).
\bibitem{bes3}
J.Z.~Bai {\it et al.}, Phys. Lett. B {\bf 589}, 7-13(2004).
\bibitem{besiii}
Internal Report, "The Preliminary Design Report of the BESIII Detector",
IHEP-BEPCII-SB-13.
\bibitem{fuko}
Y. Funakoshi, Talk at the 8th ICFA Seminar on "Future Perspectives in High
Energy Physics", Kyungpook National University, Daegu, Korea, 28
Sept - 1 Oct 2005.

\end{thebibliography}
\end{document}